\documentclass[twocolumn,showpacs]{revtex4}
\usepackage{graphicx}
\usepackage{bm}
\usepackage{color}
\usepackage{amsmath}
\usepackage{natbib}
\begin{document}

\title{The quantum effects of the spin and the Bohm potential in the oblique propagation of magnetosonic waves}

\author{Felipe A. Asenjo}
	\email{fasenjo@levlan.ciencias.uchile.cl}
\affiliation{Departamento de F\'\i sica, Facultad de Ciencias, Universidad de Chile, Casilla 653, Santiago, Chile.}
\affiliation{Departamento de Ciencias, Facultad de Artes Liberales, Universidad Adolfo Ib\'a\~nez, Diagonal Las Torres 2640, Pe\~nalol\'en, Santiago, Chile.}

\date{\today}

\begin{abstract}
We study the quantum corrections to the oblique propagation of the magnetosonic waves in a warm quantum magnetoplasma composed by mobile ions and electrons.  We use a fluid formalism to include quantum corrections due to the Bohm potential and to magnetization energy of electrons due to its spin. The effects of both quantum corrections are shown in the dispersion relation for perpendicular, parallel and oblique propagation. We find that the quantum contributions to the low-frequency depends on the type of propagation we are studying.
 
\end{abstract}

\pacs{52.25.Xz, 52.30.Ex, 52.35.-g}
\keywords{Magnetosonic waves; spin quantum plasmas.}
\maketitle

\section{Introduction}

In recent years there have been a huge interest in quantum plasmas because they can be important in different scenarios, for example in astrophysical systems \cite{baring} or high-energy lasers \cite{kremp}. Typically, for a plasma with number density $n$, the quantum effects are relevant when the thermal de Broglie wavelength $\lambda_B$ of the plasma constituents is similar to or larger than the average interparticle distance $n^{-1/3}$, i.e. when $n\lambda_B^3\geq1$.

Fluid plasma models have been developed to introduce the quantum effects of its constituents \cite{haas,marklund,brodin}. In these models, the equation of motion for particles has two main quantum corrections. One is a quantum force produced by density fluctuations, which has its origin in the so-called Bohm potential \cite{haas}. The another one is owing to the spin of particles which is considered in the equation of motion through the magnetization energy. Besides, the spin has its own equation which contains a classical evolution plus a quantum correction to its dynamics. Using this new formalism, quantum multi-fluid and quantum magnetohydrodynamic models have been proposed \cite{marklund,brodin}.

Propagation of different waves have been studied using these fluid models. Quantum corrections have been found for classical dispersion relations, either with only Bohm potential or with Bohm and spin contributions. For example, it has been studied linear waves \cite{shuk,ren}, low-frequency waves \cite{shuk3,saleem,haas2}, wave modes for dusty plasmas \cite{misra,zama}, nonlinear waves \cite{ali,shuk2}, shock waves \cite{misra2,shah} and solitons \cite{mark2}.

The quantum effects on the oblique propagation of magnetosonic waves have been studied previously for electron-positron-ion and dust-electron-ion plasmas \cite{masood}, where only the force due to the Bohm potential was included but the spin effect was not considered. In this manuscript, we study the quantum effects introduced by both the Bohm potential and the spin of electrons in the oblique propagation of low-frequency magnetosonic waves in a warm magnetoplasma composed by mobile ions and electrons. An angle $\theta$ measures how oblique is the direction of propagation of the wave with respect to a background magnetic field. The aim of our analysis is to investigate how quantum effects change for the different types of low-frequency propagation.

\section{Magnetosonic propagation in quantum plasmas}

We consider a two-fluid plasma model composed by mobile ions and electrons. This model follows the formalism of Ref.~\cite{marklund,brodin} for spin quantum plasmas. Thus, the electron specie contains both quantum effects, Bohm potential and spin force, while for the ions we neglect the quantum contributions due to their inertia. 
In our system, we use the subindex $i$ or $e$ to identify the ion or electron species. We assume that the amplitude of oscillation are small, thus we can solve the system using linearized equations.
The plasma equilibrium is assumed to have 
null zeroth order velocities for both species, ${\bf v}_{i0}=0={\bf v}_{e0}$. Thus, at first order, the continuity equations for ions and electrons are
\begin{equation}
 \frac{\partial n_i}{\partial t}+n_0 \nabla\cdot {\bf v}_i=0\, ,\qquad \frac{\partial n_e}{\partial t}+n_0 \nabla\cdot {\bf v}_e=0\, ,
\label{continuidad1}
\end{equation}
respectively. $n_0$ is the equal equilibrium number density for ions and electrons which allows quasi-neutrality. $n_i$ and $n_e$ are the first order number density perturbations of ions and electrons respectively, ${\bf v}_i$ and ${\bf v}_e$ are the first order velocity perturbations of ions and electrons respectively.

The plasma is embedded in a constant background magnetic field ${\bf B}_0$. 
Thus, the equation of motion for ions at first order is
\begin{equation}
 m_i n_0 \frac{\partial {\bf v}_i}{\partial t}=e n_0\left({\bf E}_1+\frac{{\bf v}_i}{c}\times{\bf B}_0\right)-\gamma_i k_B T_i \nabla n_i\, ,
\label{eqmotionion1}
\end{equation}
where $m_i$ is the ion mass, $e$ is the electron charge, $\gamma_i$ is the polytropic index for the ion fluid, $T_i$ is the ion fluid temperature, $k_B$ is the Boltzmann constant and $c$ is the speed of light. The pressure is related the density through the equation of state $p= k_B T n$.

The first order equation of motion for electrons is 
\begin{eqnarray}
 m_e n_0 \frac{\partial {\bf v}_e}{\partial t}&=&-e n_0\left({\bf E}_1+\frac{{\bf v}_e}{c}\times{\bf B}_0\right)-\gamma_e k_B T_e\nabla n_e\nonumber\\
&&+\frac{\hbar^2}{4m_e}\nabla (\nabla^2 n_e)-\frac{2n_0\mu}{\hbar}  \nabla({\bf s}\cdot {\bf B}_1)\, ,
\label{eqmotionelec1}
\end{eqnarray}
where $m_e$ is the electron mass, $\hbar$ is the reduced Planck constant, $\gamma_e$ is the polytropic index for the electron fluid and $T_e$ is the electron fluid temperature (for a fermionic plasma, the temperature $T_e$ is replaced by the Fermi electron temperature $T_F= \hbar^2(3\pi^2 n_0)^{2/3}/(2 m_e k_B)$ with $\gamma_e=2$), $\mu=e\hbar/(2 m_e c)$ is the electron magnetic moment and the spin ${\bf s}$ is considered as a constant.
The third and fourth terms at the right-hand side of Eq.\eqref{eqmotionelec1} correspond to the first order quantum corrections to the equation of motion of electrons \cite{marklund,brodin}. The third term is the Bohm potential and the fourth term is the spin correction to the fluid dynamics. This last term is associated to the magnetization energy of the electrons due to the electron 1/2-spin effect \cite{marklund,brodin,shuk}. 

The electromagnetic wave is governed by the Maxwell equations
\begin{equation}
 \nabla\times{\bf E}_1=-\frac{1}{c}\frac{\partial {\bf B}_1}{\partial t}\, ,
\label{ecMax1}
\end{equation}
\begin{equation}
 \nabla\times{\bf B}_1=\frac{1}{c}\frac{\partial {\bf E}_1}{\partial t}+\frac{4\pi}{c}{\bf J}\, ,
\label{ecMax2}
\end{equation}
where ${\bf J}=e n_0\left({\bf v}_i-{\bf v}_e\right)+c{\bf J}_M$ is the total current density, and ${\bf J}_M=\nabla\times(2n_e \mu {\bf s}/\hbar)$ is the magnetization spin current \cite{marklund}. This current is necessary for that the Maxwell equations take in account the spin sources of the plasma constituents.

Now, we choose the external magnetic field as ${\bf B}_0=B_0\left(\cos\theta\hat y+\sin\theta\hat z\right)$ with respect of the propagation direction determined by the wavenumber ${\bf k}=k\hat y$ of the wave. The spin of the electrons is ${\bf s}=-\left(\hbar/2\right) \eta(\alpha) \left(\cos\theta\hat y+\sin\theta\hat z\right)$, antiparallel to the background magnetic field. This minimizes the magnetic moment energy. The function $\eta(x)\equiv\tanh(x)$  is the Brillouin function due to the magnetization of a spin distribution in thermodynamic equilibrium with $\alpha= \mu B_0/(k_B T_e)$. This function appears as the solution of the spin evolution equation for spin quantum plasmas where the spin inertia and the spin-thermal coupling terms are neglected \cite{brodin}.

The first order electromagnetic fields are ${\bf E}_1=E \hat x$ and ${\bf B}_1=B \hat z=-c k E/\omega \hat z$. 
Assuming that the first order perturbations are proportional to $\exp[ik y-i\omega t]$ (where $\omega$ is the wave frequency), then the continuity equations \eqref{continuidad1} for ions and electrons can be linearized as
\begin{equation}
n_i=\frac{n_0 k}{\omega} \hat y \cdot {\bf v}_{i}\, ,\qquad n_e=\frac{n_0 k}{\omega}\hat y \cdot {\bf v}_{e}\, .
\label{continuidad2}
\end{equation}

In the same way, the linearized equations of motion \eqref{eqmotionion1} and \eqref{eqmotionelec1} becomes
\begin{equation}
-i\omega m_i n_0 {\bf v}_i=e n_0\left(E\hat x+\frac{{\bf v}_{i}}{c}\times{\bf B}_0\right)-i\gamma_i k_B T_i n_i k\hat y\, ,
\label{eqmotionion2}
\end{equation}
\begin{eqnarray}
-i\omega m_e n_0 {\bf v}_e&=&-e n_0\left( E\hat x+\frac{{\bf v}_e}{c}\times{\bf B}_0\right)-i \gamma_e k_B T_e n_ek\hat y\nonumber\\
&&-i\frac{\hbar^2}{4m_e}k^3 n_e\hat y +i\eta(\alpha)\mu n_0 B k\sin\theta\hat y\, .
\label{eqmotionelec2}
\end{eqnarray}

The general dispersion relation can be found solving the linearized version of Eqs. \eqref{continuidad2}--\eqref{eqmotionelec2} and Maxwell equations \eqref{ecMax1} and \eqref{ecMax2}, using the magnetization spin current is ${\bf J}_M=-i\mu \eta(\alpha)n_0 k^2 v_{ey} \sin\theta /\omega$.

The above equations can be solved to find the dispersion relation 
\begin{widetext}
 \begin{eqnarray}
\omega^2-c^2 k^2&=&\Omega_p^2\left[\frac{\omega^2(1-A_i)}{\omega^2(1-A_i)-\Omega_c^2+\Omega_c^2\cos^2\theta A_i}\right]\nonumber\\
&&+\omega_p^2\left[1+\frac{2\omega_c\eta(\alpha)\mu c k^2 \sin^2\theta }{e\omega^2(1-A_e)}+\frac{\eta(\alpha)^2\mu^2 c^2 k^4 \sin^2\theta}{e^2\omega^2(1-A_e)}\left(1-\frac{\omega_c^2}{\omega^2}\cos^2\theta\right)\right]\left[\frac{\omega^2(1-A_e)}{\omega^2(1-A_e)-\omega_c^2+\omega_c^2\cos^2\theta A_e}\right]\, ,\nonumber\\
&&
\label{relacdispersion1}
\end{eqnarray}
\end{widetext}
where $\Omega_p=(4\pi e^2 n_0/m_i)^{1/2}$ and $\omega_{p}=(4\pi e^2 n_0/m_e)^{1/2}$ are the ion and electron plasma frequency respectively, $\Omega_c=e B_0/(m_i c)$ and $\omega_c=e B_0/(m_e c)$ are the ion and electron cyclotron frequency respectively. We have defined the adimensional quantities 
\begin{equation}
 A_i=\frac{v_i^2 k^2}{\omega^2}\, ,\quad  A_e=\left(v_e^2+\frac{\hbar^2 k^2}{4m_e^2}\right)\frac{k^2}{\omega^2}\, ,
\label{Aparameters}
\end{equation}
where $v_i=\left(\gamma_i k_B T_i/m_i\right)^{1/2}$  and $v_e=\left(\gamma_e k_B T_e/m_e\right)^{1/2}$.

The dispersion relation \eqref{relacdispersion1} is general for any angle. The quantum corrections appear due to the spin effects, which are proportional to $\mu$, and due to the Bohm potential effect through the correction of $\hbar^2$ order in the definition of $A_e$.

From this result, we derive the low-frequency modes with the aim of to understand how the spin and the Bohm potential affect the propagation modes at different angles.

\subsection{Perpendicular propagation ($\theta=\pi/2$)}

From the complete electromagnetic propagation mode \eqref{relacdispersion1}, the magnetosonic wave is obtained as the low-frequency electromagnetic wave which propagates perpendicular to the background magnetic field (${\bf B}_0=B_0\hat z$). In this case, it appears the ${\bf  B}_0\times{\bf E}_1$ drift along the propagation direction due to the interaction of the electric field and the background magnetic field. It causes that the plasma will be compressed and released during the oscillations \cite{chen}. And, therefore, the pressure gradient force term in the equation of motion are important in this mode.

To study the low-frequency mode, we take the limit of small electron mass such that $\omega^2\ll\omega_c^2$. When we treat with classical plasmas we require that $\omega^2\ll (\gamma_e k_B T_e/m_e)k^2$, however when the quantum effects are considered, the low-frequency limit is obtained when $\omega^2\ll\omega^2 A_e$ owing to the $\hbar^2$ term in $A_e$ can be important for large $k$. Assuming besides that $\omega^2\ll\Omega_c^2$, such that $1-A_i$ can be neglected relative to $\Omega_c^2/\omega^2$, the dispersion relation \eqref{relacdispersion1} becomes
\begin{equation}
\omega^2=\frac{c^2 k^2}{v_A^2+c^2}\left({\tilde v}_A^2+v_s^2+\frac{\hbar^2 k^2}{4 m_i m_e}\left(1-\eta(\alpha)^2\right)\right)\, ,
\label{relacdispersion2}
\end{equation}
where $v_A=B_0/(4\pi n_0 m_i)^{1/2}$ is the Alfv\'en velocity, $v_s=\left(v_i^2+m_e v_e^2/m_i\right)^{1/2}$ is the acoustic speed, and ${\tilde v}_A$ is a spin-modified Alfv\'en velocity 
\begin{equation}
 {\tilde v}_A=v_A\left(1-\frac{8\pi n_0}{B_0}\eta(\alpha)\mu\right)^{1/2}\, .
\label{spinmodifv}
\end{equation}

The dispersion relation \eqref{relacdispersion2} is for a simple magnetosonic wave which propagates perpendicular to an external magnetic field in a quantum plasma. The dispersion relation has two quantum corrections. The first correction in \eqref{relacdispersion2} is due to the spin. This is a constant correction to the Alfv\'en velocity where the effect of the spin of electrons produces a contribution to the linear part of the magnetosonic mode defining the spin-modified Alfv\'en velocity \eqref{spinmodifv}.
The correction to the Alfv\'en velocity \eqref{spinmodifv} can be relevant for very dense plasmas. For astrophysical plasmas such as pulsar magnetospheres with $n_0\simeq 10^{30}\mbox{cm}^{-3}$, $B_0\simeq 10^{14}\mbox{G}$ and temperature of $10^9 \mbox{K}$ \cite{lai}, then $8\pi n_0\eta(\alpha)\mu/B_0\simeq 10^{-3}$, which is a reduction  in $0.1$\% of the Alfv\'en velocity.
On the other hand, the other quantum correction is owing to the Bohm potential and to the spin. This is a no constant correction that produces a contribution of $k^4$ order, and it depends on the both ion and electron masses. Classically the magnetosonic mode is a lineal mode, but the quantum corrections produces a nonlinear contribution that can be important for large $k$. Asociated to this correction appears the effect of the spin through the term $1-\eta^2$ which is always positive. Thus, the spin decreases the contribution of the Bohm potential to the mode.

To this quantum correction be relevant in a quantum magnetoplasma, $k$ must be of the order $k\gtrsim (m_e/n_0)^{1/2}B_0/\hbar$. For example, $k \gtrsim 10^{12}\mbox{cm}^{-1}$ for the same previous values for the density and for the magnetic field, $n_0\simeq 10^{30}\mbox{cm}^{-3}$ and $B_0\simeq 10^{14}\mbox{G}$.

\begin{figure}[!h]
\includegraphics[height=5cm]{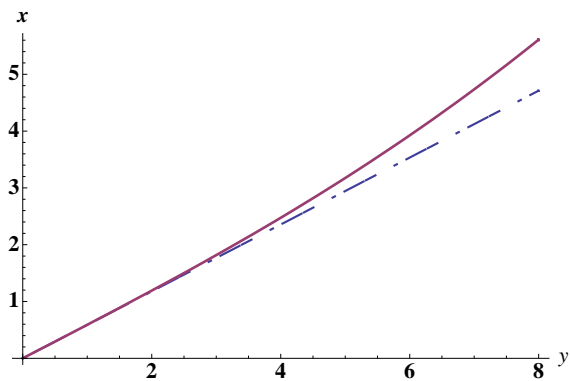}
\caption{\label{figura1} Dispersion relation \eqref{relacdispersion2} in full line, where $x=10^{-22}\omega$ and $y=10^{-22}ck$, with the frequency $\omega$ and the wavenumber $k$. The dashed line is the classical relation dispersion.  We have used $B_0\simeq10^{14}\mbox{G}$, $n_0\simeq 10^{30} \mbox{cm}^{-3}$ and $T_i=10^9\mbox{K}$.}
\end{figure}

When there are no background magnetic field, the effect of spin vanish, but the quantum correction due to the Bohm potential will not vanish because it is independent to the coupling to the magnetic field. On the other hand, classical results are recovered when $\hbar\to 0$, obtaining the well-known relation dispersion for classical magnetosonic modes \cite{chen}. The classical and the quantum corrected modes are shown in the Fig.~\ref{figura1}, where we plot the dispersion relation \eqref{relacdispersion2} in full line to compare its behavior with the classical dispersion relation [$\hbar\to 0$ in Ec.~\eqref{relacdispersion2}] which is in dashed line. The values of magnetic field and number density is choosen for some magnetized dense astrophysical plasma. These are $B_0\simeq10^{14}\mbox{G}$ and $n_0\simeq 10^{30} \mbox{cm}^{-3}$. The Bohm potential becomes important for $ck\simeq 2\cdot10^{22}\mbox{s}^{-1}$, where the dispersion relation start to show a cuadratic behavior.

The most remarkable effects of quantum contribution is the increase in the effective group velocity $\partial\omega/\partial k$ of this mode. This happen because the term proportional to $k^4$ in \eqref{relacdispersion2} whose origin is the Bohm potential. The implications of this quantum behavior in possible anomalous disperion regions are under study. Besides, this effect in the phase and group velocities can lead to  very exciting implications. For example, similar dispersion relations have been obtained for Bose-Einstein condensates where it is shown the existence of analogue black holes due to this quantum effect \cite{unruh}.

It is interesting to mention that when $\mu B_0\ll k_B T_e$, which is fulfilled for most plasmas, then we can approximate $\eta(\alpha)\approx\alpha$. Thus, the spin-modified Alfv\'en velocity \eqref{spinmodifv} becomes in $ {\tilde v}_A=v_A\left(1-\hbar^2 \omega_{p}^2/(2 k_B T_e m_e c^2)\right)^{1/2}$, which coincides with the definition for the spin-modified Alfv\'en velocity given in Ref.~\cite{brodin}. In this case,
the quantum correction can be relevant if the number density is of order $n_0\sim (m_e c^2/e^2)^3$. On the other hand, when $\mu B_0\gg k_B T_e$, then $\eta(\alpha)\approx 1$, and the effect of the Bohm potential vanish leaving the spin-modified Alfv\'en velocity $\tilde v_A^2=v_A^2(1-8\pi n_0\mu/B_0)$ as the only quantum correction to the magnetosonic wave.

\subsection{Parallel propagation ($\theta=0$)}

When the propation of the wave is parallel to the background magnetic field (${\bf B}_0=B_0\hat y$), the low-frequency limit is taken from the dispersion relation \eqref{relacdispersion1} with $\theta=0$. In this case, the spin contributions vanish and the Bohm potential corrections are simplified from the dispersion relation. 

The low-frequency mode is obtained when $\omega^2\ll\Omega_c^2$, $\omega^2\ll\omega_c^2$, $\omega^2\ll c^2k^2$ and $m_e\ll m_i$. Under this limit, the dispersion relation will be $\omega=v_A k$. This is the same result than for a classical Alfv\'en propagation mode. This is because of that the spin will be parallel to the background magnetic field and, therefore it does not couple with the perturbed magnetic field.

\subsection{Oblique propagation ($\theta=\pi/4$)}

In general, the low-frequency limit of the electromagnetic wave \eqref{relacdispersion1} can be obtained for any angle. As illustration, we consider the case of oblique propagation with $\theta=\pi/4$. For this case, we do again $\omega^2\ll\Omega_c^2$ and $\omega^2\ll\omega_c^2$. After a straightforward algebra and assuming that $\omega^2\ll c^2k^2$ and $\omega^2\ll \omega^2A_e$, we finally obtain a complicated dispersion relation for this low-frequency mode
\begin{eqnarray}
&&\frac{c^2 k^2}{\omega^2}\left(1-\frac{2\pi n_0 \mu^2 \eta(\alpha)^2}{m_e v_e^2+\hbar^2 k^2/4 m_e}\right)=\nonumber\\
&&\left(\frac{\omega^2-v_i^2k^2}{\omega^2-v_i^2k^2/2}\right)\frac{c^2}{v_A^2}-\frac{8\pi n_0 c^2\mu\eta(\alpha)}{B_0(v_e^2+\hbar^2 k^2/4 m_e^2)}\, .
\label{reldisppi4}\end{eqnarray}

For this oblique propagation, the temperature of the ion fluid and of the electron fluid play an interesting role in the dispersion relation. The ion temperature appears modifying in some sense the effective value of the Aflv\'en velocity. On the other hand, the Bohm potential is coupled to the spin in the sense that if the spin is neglected, then the Bohm potential term does not appear. Besides, the electron temperature is related with the spin and the Bohm potential. These are the main differences between this mode and the perpendicular propagation where the quantum contributions are independent of each other and both effects are independent from the temperature.  

\begin{figure}[!h]
\includegraphics[height=5cm]{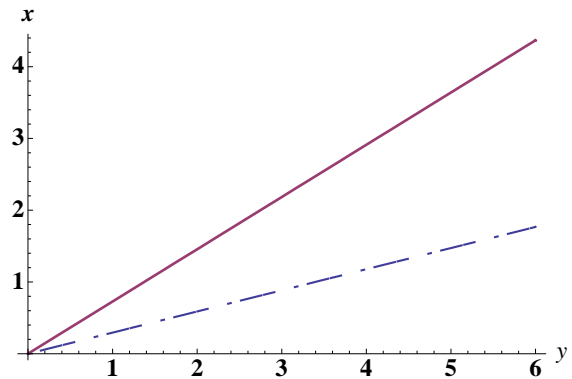}
\caption{\label{figura2} Dispersion relation \eqref{reldisppi4} in full line, where $x=10^{-22}\omega$ and $y=10^{-22}ck$. The dashed line is the classical relation dispersion $\omega^2=v_A^2k^2/2$.  We have used $B_0\simeq10^{14}\mbox{G}$, $n_0\simeq 10^{30} \mbox{cm}^{-3}$ and $T_i=10^9\mbox{K}$.}
\end{figure}

When both quantum effects, spin and Bohm potential, are ignored, we can recover the dispersion relation $\omega^2=v_A^2k^2/2$ in the limit $\omega^2\ll v_i^2 k^2$. The quantum and the classical dispersion relations are plotted in Fig.~\ref{figura2} in full and dashed lines respectively. We have used the same previous values for the density and for the background magnetic field. 

As in the case for perpendicular propagation, the quantum effects produce an increasing in the effective group velocity of the oblique linear mode. But this increasing is not enough to produce an anomalous dispersion effect, because this group velocity is less than $c$.

\section{Summary}

We have derived the quantum corrections to the propagation of low-frequency magnetosonic waves composed of mobile electrons and ions. These corrections are due to Bohm potential and to spin of electrons, both of $\hbar^2$ order. It was shown how contribution of both quantum effects change depending on the angle of the oblique propagation. 

For perpendicular propagation, the quantum correction associated to the Bohm potential introduces a non linear term which appears as an contribution of $k^4$ order in the dispersion relation. The spin of electrons produces a contribution to the Alv\'en velocity which is a correction to the linear part of the dispersion relation. 

The quantum effects does not affect the parallel propagation at low-frequency. Otherwise, for every other angle, quantum effects shown a complicated coupled to the thermal effects. In particular for $\theta=\pi/4$, the quantum contribution increase the value of the group velocity. In fact, this increasing is obtained for any angle $\theta\neq 0$. For example, for a very small angle $\epsilon$, then the dispersion relation becomes $\omega^2\approx v_A^2 k^2\left[1+2\eta(\alpha)\mu B_0\epsilon^2/(m_i v_e^2)\right]$, showing that the spin can introduces quantum correction to the propagation even at small oblique angles. 

In most plasmas, these quantum corrections are very small. However, they can be relevant for dense astrophysical scenarios with strong magnetic fields.

\begin{acknowledgments}

The author is grateful to Makarena Estrella for her support to this work.

\end{acknowledgments}

\end{document}